\documentclass[11pt]{article}

\usepackage[body={16cm,230mm}]{geometry}
\usepackage{epsfig,amsfonts,amssymb}
\usepackage{amsthm}
\usepackage{enumerate}
\usepackage{cite}
\def\g{\gamma}

\def\vt{\vartheta}
\def\beq{\begin{equation}}
\def\eeq{\end{equation}}
\def\be{\begin{displaymath}}
\def\ee{\end{displaymath}}
\def\bea{\begin{eqnarray}}
\def\eea{\end{eqnarray}}
\def\ov{\overline}
\def\bmat{\left(\begin{array}}

\def\ds{\displaystyle}

\def\Wp{ \raise.4ex\hbox{\textrm{\Large $\wp$}}}
\def\W{\mathsf{W}}

\def\f{\Phi}

\def\im{\mathop{\hbox{\rm Im}}\nolimits}

\def\TJ{{\mathsf \Theta}}
\def\HJ{{\mathsf H}}

\def\K{{\mathsf K}}
\def\Q{{\bf Q}}

\def\P{{\mathcal P}}
\def\q{{\mathsf q}}

\def\eqref#1{(\ref{#1})}
\def\?{(?)\marginpar{|?}}

\newtheorem{conj}{Conjecture}
\newtheorem*{conjA}{Conjecture A}
\newtheorem*{conjB}{Conjecture B}
\newtheorem*{conjC}{Conjecture C}
\newtheorem*{conjD}{Conjecture D}
\newtheorem*{conjE}{Conjecture E}

\renewcommand{\author}[1]{\large\rm #1\\ \bigskip}
\newcommand{\address}[1]{{\normalsize\it #1\\}\bigskip}
\renewcommand{\title}[1]{\bigskip\bigskip\Large\bf #1\bigskip\bigskip\\}

\newcounter{app}
\newcounter{sapp}[app]
\def\theapp{\Alph{app}}
\newcommand{\app}[1]{
\refstepcounter{app}{\vspace{7mm}
\noindent\Large\bf Appendix
\theapp.
 \ #1 \par \vspace{5mm}}
\setcounter{equation}{0}
\def\theequation{\Alph{app}.\arabic{equation}}}

\begin{document}

%\begin{flushright}
%{\sf version 1.0 \it (5/10/2004)}
%\end{flushright}

\vglue .3 cm

\begin{center}
\title{Eight-vertex model and Painlev\'e VI equation\\
  II. Eigenvector results}

\author{         Vladimir V. Mangazeev\footnote[1]{email:
                {\tt vvm105@rsphysse.anu.edu.au}}and 
                Vladimir V. Bazhanov\footnote[2]{email:
                {\tt Vladimir.Bazhanov@anu.edu.au}}}

\address{Department of Theoretical Physics,\\
         Research School of Physical Sciences and Engineering,\\
    Australian National University, Canberra, ACT 0200, Australia.}

\end{center}

\setcounter{footnote}{0}
\vspace{5mm}

\begin{abstract}
We study 
a special anisotropic $\mathsf{XYZ}$-model on a periodic chain of an
odd length and conjecture exact expressions for certain components of the
ground state eigenvectors. The results are written in terms of tau-functions
associated with Picard's elliptic solutions of the Painlev\'e VI equation.  
Connections with other problems related to the eight-vertex model are
briefly discussed.  
\end{abstract}

\newpage
%%%%%%%%%%%%%%%%%%%%%%%%%%%%%%%%%%%%%%%%%%%%%%%%%%
\section{Introduction}\label{sect:intro}
This is a sequel to our paper \cite{BM06a} devoted to connections of
the eight-vertex model of statistical mechanics \cite{Bax72} 
with the theory of Painlev\'e transcendents.  
Here we study a related case of the anisotropic $\mathsf{XYZ}$-model  
on a periodic chain of an odd length, $N=2n+1$. 
 
Let $\sigma^{(j)}_x$, $\sigma^{(j)}_y$ and $\sigma^{(j)}_z$,
$j=1,\ldots,N$,\  denote 
usual Pauli matrices acting at the $j$-th site of the chain.
Consider a particular $\mathsf{XYZ}$-Hamiltonian
\beq
{\bf H}_{\mathsf{XYZ}}=-\frac{1}{2}\, \sum_{j=1}^N
\big(J_x\,\sigma^{(j)}_x\sigma^{(j+1)}_x+
J_y\,\sigma^{(j)}_y\sigma^{(j+1)}_y+
J_z\,\sigma^{(j)}_z\sigma^{(j+1)}_z\big),\label{ham}
\eeq
where the coefficients 
\beq
J_x=\frac{2\,(1+\zeta)}{\zeta^2+3},\quad
J_y=\frac{2\,(1-\zeta)}{\zeta^2+3},\quad 
J_z=\frac{\zeta^2-1}{\zeta^2+3},\label{J-def}
\eeq
are specific rational functions of a single parameter $\zeta$, 
which satisfy the relation 
\beq
J_x \,J_y + J_y \,J_z +J_z\, J_x=0\ .\label{constr}
\eeq
Baxter \cite{Bax72b} proved that for an
infinitely large chain, $N\to\infty$, 
the ground state eigenvalue of \eqref{ham}
in this case has a very simple form      
\beq
\frac{E_0(\zeta)}{N}=-\frac{J_x+J_y+J_z}{2} =-\frac{1}{2}\, .\label{E0}
\eeq
Later on it was conjectured  \cite{Str01a} that this expression 
is exact for all finite odd values of $N$. 

The Hamiltonian \eqref{ham}
commutes with the transfer matrix of the eight-vertex model (8V-model),
where the Boltzmann weights $a$, $b$, $c$, $d$ (we use standard notations of
\cite{Bax72}, see Sect.~\ref{eight} below for further details) are
constrained as  
\beq
(a^2+ab)\,(b^2+ab)=(c^2+ab)\,(d^2+ab)\ ,\label{abcd-rel}
\eeq
and the variable $\zeta$ in \eqref{J-def}  is given by 
\beq
\zeta=\frac{cd}{ab},
\qquad 
\g=
\frac{(a-b+c-d)(a-b-c+d)}{(a+b+c+d)(a+b-c-d)}. 
\label{zeta-def}
\eeq
Additional variable $\gamma$, introduced for later convenience, 
is connected to $\zeta$ by a 
simple self-reciprocal rational substitution 
\beq
\zeta=\frac{\gamma+3}{\gamma-1},\qquad
\gamma=\frac{\zeta+3}{\zeta-1}\ .\label{zega}
\eeq

The spectrum of \eqref{ham} possesses an $S_3$ symmetry group with
respect to permutations of the constants $J_x$, $J_y$ and
$J_z$. Indeed, 
any such permutation can be compensated by a linear
transformation which acts on the eigenvectors 
and does not affect the spectrum. 
For the parametrization
\eqref{J-def} this group is
generated by two substitutions of the variable $\zeta$, 
\beq\begin{array}{llll}
{\mathbf s}_{\mathsf{xy}}:&\quad \zeta\to-\zeta\quad&\Longrightarrow\ 
&J_x\leftrightarrow J_y,\  J_z\to J_z,\\[.2cm]
{\mathbf s}_{\mathsf{xz}}:&\quad \zeta\to\frac{\zeta+3}{\zeta-1}\quad
&\Longrightarrow\ 
&J_x\leftrightarrow J_z,\  J_y\to J_y\ .
\end{array}\label{s3group}
\eeq

The largest eigenvalue of the transfer matrix, corresponding to
\eqref{E0}, also has a remarkably simple conjectured form  \cite{Str01a}  
\beq
\Lambda_0=(a+b)^N, \qquad N=2n+1, \label{tab}
\eeq
which is expected to hold for finite chains. 

In \cite{BM05} we studied Baxter's famous ${\mathsf {TQ}}$-equation 
for this simple eigenvalue \eqref{tab} and found corresponding
eigenvalues of the $\mathsf{Q}$-operator. 
With an appropriate normalization they can be expressed through
certain polynomials  
\beq
\P_n(x,z)=\sum_{k=0}^n r^{(n)}_k(z)\,x^k,\qquad z=\gamma^{-2},
\qquad n=0,1,2,\ldots\ ,\label{P-def0}
\eeq
of the variable $x$, defined by the following quadratic equation (see
also \eqref{x-def1}), 
\beq
\Big(\sqrt{x}-\frac{\g}{\sqrt{x}}\Big)^2=-\frac{16\,(a-b)^2\,c\,d}{(c+d)^2\,
(a+b+c+d)(a+b-c-d)}.\qquad \label{x-def}
\eeq
The coefficients $r^{(n)}_i(z)$, $i=0,\ldots,n$,  appearing in
\eqref{P-def0}, are 
polynomials in the variable 
$z=\gamma^{-2}$ with positive integer coefficients. 
Detailed definitions of the polynomials \eqref{P-def0} 
are presented in Sect.~\ref{eight}. Here we want to illustrate their 
connection to the Painlev\'e VI equation. This connection manifests itself in
some specific properties of the coefficients  $r^{(n)}_i(z)$. In
particular, let $s_n(z)\equiv r^{(n)}_n(z)$ be a coefficient in front
of the leading power of $x$ in \eqref{P-def0}.
In \cite{BM05} we conjectured the following recurrence relation
\beq
\begin{array}{ll}
2z(z-1)(9z-1)^2\,\partial_z^2\,\log s_n(z)+2(3z-1)^2(9z-1)\partial_z
\log s_n(z)+&\label{gen19}\\[.2cm]
\>\>\>+8(2n+1)^2{\ds\frac{s_{n+1}(z)s_{n-1}(z)}{s_n^2(z)}}-
[4(3n+1)(3n+2)+(9z-1)n(5n+3)]=0,\nonumber
\end{array}
\eeq
where $s_0(z)=s_1(z)\equiv 1$, which uniquely
determines  the polynomials $s_n(z)$, for all $n\in{\mathbb Z}$.
%$n=0,\pm1,\pm2\ldots$. 
Later on 
we proved \cite{BM06a} that Eq.\eqref{gen19} exactly coincides
with the recurrence relation for the tau-functions 
associated with special elliptic solutions of the Painlev\'e VI equation. 
In this letter we extend these connections to study 
{\em ground state eigenvectors} of ${\bf H}_\mathsf{XYZ}$, 
corresponding to the eigenvalue \eqref{E0}. 

For odd $N$ all eigenvalues of
\eqref{ham} are double degenerate. Thus, there are two ground state
eigenvectors 
\beq
{\bf H}_{\mathsf{XYZ}}\,\Psi_\pm = E_0\, \Psi_\pm,\qquad {\mathcal S}\,
\Psi_\pm=\pm\Psi_\pm,\qquad {\mathcal R}\,\Psi_\pm=\Psi_\mp\ ,\label{Psi-def}
\eeq
where 
\beq
{\mathcal S}=
\sigma^{(1)}_z\otimes\sigma^{(2)}_z\otimes\cdots\otimes\sigma^{(N)}_z,\qquad
{\mathcal R}=
\sigma^{(1)}_x\otimes\sigma^{(2)}_x\otimes\cdots\otimes\sigma^{(N)}_x\ ,
\label{rs-def}
\eeq
and 
\beq
\big[\,{\bf H}_{\mathsf{XYZ}},\,{\mathcal S}\,\big]=
\big[\,{\bf H}_{\mathsf{XYZ}},\,{\mathcal R}\,\big]=0,\qquad
      {\mathcal R} \, {\mathcal S}=(-1)^N\, {\mathcal S}\, {\mathcal
        R}\ .
\eeq     
Due to the spin reversal symmetry, generated by the operator ${\mathcal
  R}$, it is enough to consider one of these vectors. For
definiteness, consider the
vector 
$\Psi_-$. Omitting the suffix ``$-$'', we denote its components as
$\Psi_{i_1,i_2,\ldots,i_N}$, where $i_1,i_2,\ldots,i_N\in \{0,1\}$ and
assume an orthonormal basis $|i\rangle$, $i=0,1$, for each spin 
\beq
\sigma_z\,|0\rangle=+|0\rangle,\qquad
\sigma_z\,|1\rangle=-|1\rangle\ .
\eeq
The ground state eigenvectors are translationally invariant
and possess a left-right reflection symmetry. Taking this into account we will
give only one non-vanishing representative from each symmetry class. 
Note that 
for non-vanishing components
$\Psi_{i_1,i_2,\ldots,i_N}$ of $\Psi_-$ the 
number of ``down-spins'' in the set $\{i_1,i_2,\ldots,i_N\}$ is 
odd, while for vanishing components it is even,
\beq
\Psi_{i_1,i_2,\ldots,i_N}\equiv 0, \qquad \mbox{if}\qquad
  i_1+i_2+\cdots+i_N=0\pmod 2 \ .
\eeq 
The fact that both the coefficients \eqref{J-def} and the
eigenvalue \eqref{E0} are {\em rational functions in $\zeta$ with
integer coefficients} implies that
with a suitable normalization one can make all components 
of the eigenvector $\Psi_-$
to be {\em polynomials in $\zeta$ with integer coefficients} \cite{Str01c}   
(such that there are no polynomial factors common for all 
components). 
This choice is unique up 
to a numerical normalization. The 
later is fixed by the requirements 
\beq
\Psi_{\scriptsize\underbrace{0,\ldots,0}_{n+1}\underbrace{1,\ldots,1}_{n}}
\Big\vert_{\zeta=0}=1, \quad \mbox{for odd $n$};\qquad\qquad
\Psi_{\scriptsize\underbrace{0,\ldots,0}_{n}\underbrace{1,\ldots,1}_{n+1}}
\Big\vert_{\zeta=0}=1, \quad \mbox{for even $n$}.\label{6vnorm}
\eeq
Note that in the case $\zeta=0$ the Hamiltonian \eqref{ham} 
reduces to that of 
the $\mathsf{XXZ}$-model with 
the parameter $\Delta=-1/2$. From this point of view the normalization
\eqref{6vnorm} is identical to that used in \cite{SR2001}, where this
particular $\mathsf{XXZ}$-model was studied.

We have calculated all components of the eigenvectors directly from the
definition \eqref{Psi-def} for $N\le 17$ (and some particular 
components for $N\le25$) and made several 
interesting observations which we formulate as conjectures valid for
all $N=2n+1$. As an example we present here
%\begin{quote}
\begin{conj}{The norm of the eigenvector $\Psi_-$ is given by 
\beq
\left\vert\Psi_-\right\vert^2=\sum_{{i_1,i_2,\ldots,i_N}\in\{0,1\}} 
\Psi^2_{{i_1,i_2,\ldots,i_N}}=(4/3)^n\, \zeta^{n(n+1)}\,
s_n(\zeta^{-2})\, s_{-n-1}(\zeta^{-2})\ ,\label{psi-norm}
\eeq
where $s_n(\zeta^{-2})$,\  $n\in{\mathbb Z}$, 
are defined by the recurrence relation
\eqref{gen19} with $z=\zeta^{-2}$ and $s_0(z)=s_1(z)\equiv1$. 
 }
\end{conj}
%\end{quote}
Other conjectures on the properties of eigenvectors 
require additional notations; they are presented in
Sect.~\ref{results}. 
Basic definitions for the 8V-model, a brief review of some
of our previous results \cite{BM05} and some new results 
on the eigenvalues of Baxter's $\mathsf{Q}$-operators  are given in 
Sect.~\ref{eight}.
In Conclusion we discuss some unresolved questions and connections 
of our results to other problems related with the eight-vertex model, in
particular, to the eight-vertex solid-on-solid model \cite{Bax73abc} 
with the domain wall boundary condition \cite{R09} and the
three-coloring problem \cite{Bax70c}.

%%%%%%%%%%%%%%%%%%%%%%%%%%%%%%%%%%%%%%%%%%%%%%%%%%
\section{The eight-vertex model and $\mathsf{TQ}$-equation}\label{eight}
%%%%%%%%%%%%%%%%%%%%%%%%%%%%%%%%%%%%%%%%%%%%%%%%%%

\subsection{Basic definitions and notations}
We consider the eight-vertex model on the $N$-column
square lattice with the periodic (cylindrical) 
boundary conditions and assume that $N$ is an odd integer $N=2n+1$.
Following \cite{Bax72} we parametrize the Boltzmann weights
$a$, $b$, $c$, $d$ of the model as\footnote{We
use the notation of \cite{WW} for theta-functions
  $\vt_k(u\,|\,\q)$, $k=1,2,3,4$, of
the periods $\pi$ and $\pi\tau$, $\q=e^{i\pi\tau}$, $\im \tau>0$.
The theta-functions $\HJ(v)$, $\TJ(v)$ of the nome $\q_B$
used in \cite{Bax72} are given by
$$
\q_B=\q^2,\quad \HJ(v)=\vt_1(\frac{\pi v}{2\K_B}\,|\,\,\q^2),\quad \TJ(v)=
\vt_4(\frac{\pi v}{2\K_B}\,|\,\,\q^2),
$$
where  $\K_B(k)$ is the complete elliptic integral of the first kind
with the elliptic modulus $k=\vt^2_2(0|\q_B)/\vt^2_3(0|\q_B)$.}
\bea
&a=\rho
  \ \vt_4(2\eta\,|\,\q^2)\ \vt_4(u-\eta\,|\,\q^2)\ \vt_1(u+\eta\,|\,\q^2),&
\nonumber\\
&b=\rho
  \ \vt_4(2\eta\,|\,\q^2)\ \vt_1(u-\eta\,|\,\q^2)\ \vt_4(u+\eta\,|\,\q^2),
&\nonumber\\
&c=\rho
  \ \vt_1(2\eta\,|\,\q^2)\ \vt_4(u-\eta\,|\,\q^2)\ \vt_4(u+\eta\,|\,\q^2),
&\label{weights}\\
&d=\rho
  \ \vt_1(2\eta\,|\,\q^2)\ \vt_1(u-\eta\,|\,\q^2)\ \vt_1(u+\eta\,|\,\q^2),
&\nonumber
\eea
and fix the normalization factor,
\beq
\rho=2\,\ \vt_2(0\,|\,\q)^{-1}\ \vt_4(0\,|\,\q^2)^{-1}.\label{ei2}
\eeq
With this parametrization the constraint \eqref{abcd-rel} is
equivalent to the condition
\beq
\eta=\pi/3 \ ,         \label{pi3}
\eeq
which will be always assumed throughout this paper. This still leaves two
arbitrary parameters: the (spectral) parameter $u$ and the elliptic nome
$\q=e^{i\pi\tau}$, $\rm{Im}\,\tau>0$.
The variables $\zeta$, $\gamma$ and $x$, defined in \eqref{zeta-def},
\eqref{zega} and \eqref{x-def}, can  be written as 
\beq
\zeta=
\biggl[\frac{\vt_1(\frac{2\pi}{3}\,|\,\q^2)}
{\vt_4(\frac{2\pi}{3}\,|\,\q^2)}\biggl]^2,
\quad 
\g=-\biggl[\frac{\vt_1(\frac{\pi}{3}\,|\,\q^{1/2})}
{\vt_2(\frac{\pi}{3}\,|\,\q^{1/2})}\biggr]^2,\quad
x=\gamma
\biggl[\frac{\vt_3(\frac{u}{2}\,|\,\q^{1/2})}
{\vt_4(\frac{u}{2}\,|\,\q^{1/2})}\biggr]^2,
\quad z=\gamma^{-2}\ . \label{x-def1}
\eeq
Note that the last expression for $x$ determines our choice of a particular  
root of the quadratic equation \eqref{x-def}.

\subsection{The $\mathsf{TQ}$-equation}\label{tq-sect} 
Any eigenvalue, $T(u)$, of the row-to-row transfer matrix\footnote{%
We use exactly the same definition of the transfer
  matrix as in \cite{Bax72}.}  of the
8V-model
satisfies Baxter's famous $\mathsf{TQ}$-equation \cite{Bax72},
\beq
T(u)\,Q(u)=\phi(u-\eta)\,Q(u+2\eta)+\phi(u+\eta)\,Q(u-2\eta),  \label{TQ}
\eeq
where, with an account of \eqref{ei2},  
\beq
\phi(u)=\vt_1^N(u\,|\,\q).\label{phi-def}
\eeq
With the parametrization \eqref{weights}, \eqref{ei2} 
the eigenvalue \eqref{tab} takes the
form 
\beq
T(u)=(a+b)^N=\phi(u),\quad \eta=\pi/3, \quad N=2n+1. \label{T-simple}
\eeq
Equation \eqref{TQ} for this eigenvalue,
 has been
studied in \cite{BM05}. 
It has two different solutions \cite{BLZ97a,KLWZ97}, denoted
$Q_\pm(u)\equiv
Q_\pm(u,\q,n)$, which are {\em entire}
functions of the variable $u$ and obey the following periodicity
conditions \cite{Bax72,McCoy2},
\beq
Q_\pm(u+\pi)=\pm (-1)^n Q_\pm(u),
\quad Q_\pm(u+\pi\tau)=\q^{-N/2}\ e^{-iNu}\ Q_\mp(u),\quad
Q_\pm(-u)=Q_\pm(u). \label{Qper}
\eeq
The above requirements uniquely determine $Q_\pm(u)$ to within 
$u$-independent normalization factors. 
The solutions $Q_\pm(u)$ satisfy
the quantum Wronskian relation \cite{BLZ97a,BM07} 
\beq
Q_+(u+\eta)\,Q_-(u-\eta)-Q_+(u-\eta)\,Q_-(u+\eta)=2i\phi(u)\,W(\q,n),\label{Wr}
\eeq
where $W(\q,n)$ is a function of $\q$ and $n$ only 
(The fact that $W(\q,n)$ does not depend on the variable $u$ 
follows from \eqref{TQ} and \eqref{Qper}). 
Note that, taking into account the periodicity \eqref{Qper}, one can
bring Eq.\eqref{TQ} to the form
\beq
\phi(u)\,Q(u)+\phi(u+2\pi/3)\,Q(u+2\pi/3)
+\phi(u+4\pi/3)\,Q(u+4\pi/3)=0\ .\label{TQ2}
\eeq
Below it will be  more convenient to use the combinations
\beq
Q_1(u)=(Q_+(u)+Q_-(u))/2,\quad Q_2(u)=
\,(Q_+(u)-Q_-(u))/2\ , \label{Q12-def}
\eeq
which are simply related by the periodicity relation
\beq
Q_{1,2}^{(n)}(u+\pi)=(-1)^n\,Q_{2,1}^{(n)}(u).\label{Qper3}
\eeq
Bearing this in mind we will only quote
results for $Q_1(u)$, writing it as $Q^{(n)}_1(u)$ to indicate the
$n$-dependence. 
Introduce new functions 
${\mathcal P}_n(u)$ instead of $Q_{1}^{(n)}(u)$,
\beq
Q^{(n)}_1(u)={\mathcal N}(\q,n)\, \vt_3(u/2\,|\,\q^{1/2})\>
\vt_4^{\>2n}(u/2\,|\,\q^{1/2})\>{\mathcal P}_n(u),
\label{Q1}
\eeq
where ${\mathcal N}(\q,n)$ is an arbitrary normalization factor. 
The analytic
properties of $\P_n(u)$ are determined by the periodicity relations
\eqref{Qper} and the fact that the eigenvalues 
$Q^{(n)}_{1,2}(u)$ are entire functions of the variable $u$. A simple
analysis shows that $\P_n(u)$ is an even doubly periodic function of $u$,
\beq
\P_n(u)=\P_n(u+2\pi)=\P_n(u+\pi\tau),\qquad \P_n(u)=\P_n(-u)\ ,
\eeq
with all its poles\footnote{%
An apparent pole at $u=\pi+\pi
\tau/2$ cancels out because 
$Q_1^{(n)}(u)$ vanishes at this point 
as a consequence of 
the periodicity conditions \eqref{Qper}.}
(of the order $2n$ and lower)
located at the point $u=\pi\tau/2$.
Every such function is an $n$-th degree   
polynomial in the variable $x$, given by \eqref{x-def1} (see \S~20.51 
of ref.~\cite{WW}). The coefficients in these polynomials will, of course, 
depend on the elliptic nome $\q$. Let us now 
change independent variables from $u$ and $\q$ to the variables $x$ and
$z=\gamma^{-2}$, defined in \eqref{x-def1}, and (with a slight abuse
of notations) write  
$\P_n(u)$ as 
\beq
{\mathcal P}_n(x,z)=\sum_{k=0}^n r^{(n)}_k(z)\,x^k.\label{P-def}
\eeq

The $\mathsf{TQ}$-equation \eqref{TQ2} can 
be re-written in terms of the polynomials ${\mathcal P}_n(x,z)$. 
For a fixed value of the nome $\q$, the variable $x$ in \eqref{x-def1} 
is a function of $u$, so we can write it as $x=x(u)$. 
Introduce two new variables 
\beq
x_\pm=x(u\pm\textstyle{\frac{\pi}{3}})=\gamma^2/x(u\mp\frac{2\pi}{3}).
\label{xpm-def}
\eeq
They satisfy the relations  
\beq
x_+x_-=\frac{(x-1)^2}{(x\, z-1)^2},\quad
x_++x_-=\frac{2\,z\,(x^2\,z+1)-x\,(z^2+4\,z-1)}{z\,(x \,z-1)^2}\ ,
\label{xpm-alg}
\eeq
which can be easily solved for $x_\pm$ in terms of $x$ and $z$. 
The resulting expressions involve a square root from a third order
polynomial in $x$.
It is convenient to define
\beq
f_\pm=\frac{1}{2}\pm
\frac{x\,(z-1)[(2\,x-3)\,z+1]}{2\,z\,(x_--x_+)\,(x\,z-1)^2},
\quad\rho_\pm=\frac{x_\pm-1}{(1-z \,x_\pm)\,x}.
\label{rho-def}
\eeq
With all these new notations the $\mathsf{TQ}$-equation \eqref{TQ2}
can now be transformed to its algebraic form,
\beq
{\mathcal P}_n(x,z)=
\rho_+\,f_-^{2n+1}\/\,
{\mathcal P}_n\big(z^{-1}x_-^{-1}\,,z\big)+
\rho_-\,f_+^{2n+1}\/\,
{\mathcal P}_n\big(z^{-1}x_+^{-1}\,,z\big)\ .
\label{TQ-alg}
\eeq

Substituting \eqref{P-def} into the last equation, 
and expanding it near the point $x=0$, one immediately
obtains a simple relation, 
\beq
r_n^{(0)}(z)\equiv\P_n(0,z)
=4^{-n}\,z^{-1}\,(z+n(3z-1))\/\,\P_n(z^{-1},z)-
4^{-n}z^{-2}\,(z-1)\/\,\frac{\partial \P_n(x,z)}{\partial
  x}\Big\vert_{x=z^{-1}}\ ,
\eeq
quoted here for future references.

\subsection{Polynomials ${\mathcal P}_n(x,z)$}
Let us now substitute polynomials
\eqref{P-def} into the 
$\mathsf{TQ}$-equation \eqref{TQ-alg}. 
Excluding $x_+$ and $x_-$ with the help of \eqref{xpm-alg},  
one can readily see that the RHS of \eqref{TQ-alg} is a rational
function of 
$x$ (indeed, it is a symmetric function of $x_+$ and
$x_-$ and, therefore, can be expressed through two elementary combinations  
\eqref{xpm-alg}). Writing Eq.\eqref{TQ-alg} as a polynomial in $x$ and
equating its coefficients to zero one obtains an (overdetermined)
system of homogeneous linear equations for $n+1$ unknowns 
$r^{(n)}_0(z),r^{(n)}_1(z),\ldots, r^{(n)}_n(z)$. All elements of the
coefficient matrix for this system are 
rational functions of the variable $z$ with integer coefficients. This
means that with a suitable normalization all $r^{(n)}_k(z)$,
$k=0,1,\ldots,n$, can be made polynomials in $z$ with integer
coefficients (such that there are no polynomial factors common for all
$r^{(n)}_k(z)$).  Thus, $\P_n(x,z)$ are two-variable polynomials in $x$ and
$z$ with integer coefficients. The first few of them are given in
\eqref{pol3} and Appendix~A below.

Originally, 
we have calculated \cite{BM05} these polynomials for
$n\le10$ by directly solving Eqs.\eqref{TQ2} and \eqref{TQ-alg} by a
combination of analytical and numerical techniques.  
For larger $n$ this did not appear be to practical 
due to  complexity of intermediate  
expressions.  Subsequently, in the same paper \cite{BM05}, 
we found a more efficient
method for the calculation of $\P_n(x,z)$, based on the partial differential
equation \eqref{P-pde}, discussed below. We have observed that all
coefficients of $\P_n(x,z)$ are, in fact, {\em positive} integers 
for all $n\le100$ and
suggested that these coefficients might have a combinatorial
interpretation (which is yet to be found). 

Below we summarize all important properties of ${\mathcal P}_n(x,z)$ 
discovered in our previous works \cite{BM05,BM06a}.

\begin{conjA}[%
%Bazhanov-Mangazeev 
\cite{BM06a,BM05}]{\ }
\begin{enumerate}[(a)]
\item
The degrees of the polynomials $r^{(n)}_i(z)$, $i=0,\ldots,n$,\ 
appearing as coefficients in the expansion \eqref{P-def}, 
are given by
\beq
\deg [r^{(n)}_k(z)]\le \big\lfloor\,{n(n-1)}/{4}+{k}/{2}\,\big\rfloor\ ,
\eeq
where 
$\lfloor x\rfloor$
denotes the largest integer not exceeding $x$.
\item
if the normalization of $\P_n(x,z)$ is
fixed by the requirement 
\beq
r^{(n)}_n(0)=1\ ,\label{r-norm}
\eeq
then all polynomials  $r^{(n)}_k(z)$, 
$k=0,1,\ldots,n$, \ have {\bf positive integers} coefficients 
in their expansions 
in powers of $z$. 

\end{enumerate}
\end{conjA}

%\noindent
The normalization \eqref{r-norm} will be
implicitly assumed throughout the rest of the paper. 
The most important property of the polynomials $\P_n(x,z)$ is
that they satisfy a remarkable linear partial differential equation. This
equation can be written in different forms, depending on the choice of
independent variables and unknown function. First consider the case
of the original variables $u$ and $\q$ of the 8V-model.
Introduce the functions 
\beq
\f_\pm(u,\q,n)=
\frac{\vt_1^{2n+1}(u\,|\,\q)}{\vt_1^n(3u\,|\,\q^3)}Q_\pm(u,\q,n),\label{fpm}
\eeq
where $Q_\pm(u,\q,n)$ are eigenvalues of the
$\mathsf{Q}$-operators, defined in Sect.\ref{tq-sect}. 
The analytic properties of $\f_\pm(u,\q,n)$ in the variable $u$ 
are determined by \eqref{Qper}.

\begin{conjB} [\cite{BM05}] The functions $\f_\pm(u,\q,n)$, defined by
  \eqref{fpm}, satisfy the 
non-stationary Schr\"odinger equation
\beq
6\,q\frac{\partial}{\partial q}\f(u,q,n)=
\Big\{-\frac{\partial^2}{\partial u^2}+
9\, n\, (n+1)\,\Wp(3u\,|\,q^3)+c(q,n)\Big\}\f(u,q,n).
\label{lame-pde}\eeq
\end{conjB}
Here the modular parameter $\tau$ plays the role of the (imaginary) time
and the time-dependent potential is defined through
the elliptic Weierstrass $\Wp$-function \cite{WW}
(our function $\Wp(v\,|\,e^{i\pi\epsilon})$ has the periods
$\pi$ and $\pi\epsilon$).
The constant $c(q,n)$  appearing in \eqref{lame-pde}
is totally controlled by the normalization of $Q_\pm(u)$ and can be
explicitly determined once this normalization is fixed (see Eqs.(37) and 
(38) in  \cite{BM05}).
Equation \eqref{lame-pde} is obviously related to the Lam\'e
differential equation
and could be naturally called the ``non-stationary Lam\'e
equation''. This equation arises in various contexts \cite{EK94,Fat09}   
which are not immediately related to this paper. 

The differential equation \eqref{lame-pde} can be equivalently rewritten in an
algebraic form for the polynomials $\P_n(x,z)$.  
\beq
\Big\{A(x,z)\,\partial_x^2 +B_n(x,z)\,\partial_x +C_n(x,z)\, +
T(x,z)\,{\ds\partial_z}\Big\}\ \P_n(x,z)=0,\label{P-pde}
\eeq
where
\beq\begin{array}{rcl}
A(x,z)&=&2x(1 + x - 3xz + x^2z)(x + 4z - 6xz -3xz^2 + 4x^2z^2),\\[.3cm]
B_n(x,z)&=&4(1+x-3xz+x^2z)(x+3z-7xz+3x^2z^2)+\\[.3cm]
&&\kern-3.0em
+2nx(1-14z+21z^2-8x^3z^3+3x^2z(3z^2+6z-1)-x(1-9z+23z^2+9z^3)),\\[.3cm]
C_n(x,z)&=&n\,\big[z(9z-5)+x^2z(3z^2+11z-2)+x(9z^3-38z^2+19z-2)-\\[.3cm]
&&\kern-1.0em
-4x^3z^3+nz(1-9z-x(9z^2-36z+3)+x^2(3z^2-31z+4)+8x^3z^2)\big],\\[.3cm]
T(x,z)&=&-2z(1-z)(1-9z)(1 + x - 3xz + x^2z).\label{gen10}
\end{array}
\eeq
%\noindent 
It is fairly easy to prove \cite{BM05} that the differential equation
\eqref{lame-pde}, restricted to a class of functions $\f(u,\q,n)$ with  
suitable analytic properties in the
variable $u$,  
implies the functional equation \eqref{TQ2}.  
The non-trivial part of the Conjecture~B is the fact of 
{\em existence} of solutions 
of \eqref{lame-pde} with these analytic properties. For
Eq.\eqref{P-pde} this translates into a question  
of existence of solutions, which are polynomials in the variable $x$.

Equation \eqref{P-pde} is 
extremely useful for finding polynomial solutions,
even though the coefficients therein look very complicated.
The first polynomials $\P_n(x,z)$ read
\beq\begin{array}{l}
\P_0(x,z)=1, \quad \P_1(x,z)=x+3,\quad
\P_2(x,z)=x^2(1+z)+5x(1+3z)+10,\\[.3cm]
\P_3(x,z)=x^3(1+3z+4z^2)+7x^2(1+5z+18z^2)+7x(3+19z+18z^2)+35+21z,
\end{array}\ ,\label{pol3}
\eeq
the next one is given in Appendix~A.
The constant term and leading coefficient in these polynomials (with
respect to the variable $x$) are determined by the following 
\begin{conjC}[\cite{BM05,BM06a}]
The coefficients for the lowest and highest powers of 
$x$ in $\P_n(x,z)$, corresponding to $k=0$ and $k=n$ in \eqref{P-def}, read
\beq
\overline{s}_n(z)\equiv r_0^{(n)}(z)=\tau_n(z,-1/3),
\quad s_n(z)\equiv r_n^{(n)}(z)=\tau_{n+1}(z,1/6)\ ,
\label{r-coeff}
\eeq
where the functions $\tau_n(z,\xi)$ (for each fixed value of the
their second argument $\xi$) are determined by the recurrence
relation
\bea
&2z(z-1)(9z-1)^2[\log \tau_n(z)]''_z+2(3z-1)^2(9z-1)[\log
  \tau_n(z)]'_z
+\nonumber\\
&
{\ds+
8\Bigl[2n-4\xi-\frac{1}{3}\Bigl]^2{\ds\frac{\tau_{n+1}(z)
\tau_{n-1}(z)}{\tau_n^2(z)}}-}\nonumber&\\
&{-[12(3n-6\xi-1)(n-2\xi)+(9z-1)(n-1)(5n-12\xi)]=0},\label{tau-rec}&
\eea
with the initial condition
\beq
\tau_0(z,\xi)=1,\quad \tau_1(z,\xi)=-4\xi+5/3\ .\label{tau-init}
\eeq
The functions $\tau_n(z,\xi)$
are polynomials in $z$
for all $n=0,1,2,\ldots,\infty$.
%\end{enumerate}
\end{conjC}
As explained in \cite{BM05}, the partial differential equation \eqref{P-pde}
leads  to a
descending recurrence relations for the coefficients in \eqref{P-def}, in
the sense that each coefficient $r^{(n)}_k(z)$ with $k<n$ can be
recursively calculated in terms of $r^{(n)}_m(z)$, with $m=k+1,\ldots,n$
and, therefore, can be eventually expressed through the coefficient
$r^{(n)}_n(z)$ of the leading power of $x$. Conditions
that this procedure truncates (and thus defines a polynomial, but not an
infinite series in negative powers of $x$) completely determine the
  starting leading coefficient as a function of $z$.  
The above conjecture implies that these truncation conditions are
  equivalent to the recurrence
  relation \eqref{gen19} which is a particular case of
  \eqref{tau-init} for $\xi=1/6$. Similar reasonings 
  apply to the coefficient $r^{(n)}_0(z)$ in \eqref{P-def} (the constant term
  with respect to the variable $x$). 

Note, that Eq.\eqref{tau-rec} exactly coincides \cite{BM06a} 
with the recurrence relation for the tau-functions 
associated with special elliptic solutions of the Painlev\'e VI
equation.

\subsection{Quantum Wronskian}

We conclude this section with a short analysis of the algebraic form 
\beq\begin{array}{l}
\ds\frac{x_-^n}{1-x+{x_-(1-xz)}}\,\P_n(x_+\,,z)\,\P_n(z^{-1} x_-^{-1}\,,z)+
\frac{x_+^n}{1-x+{x_+(1-xz)}}
\,\P_n(x_-,\,z)\,\P_n(z^{-1} x_+^{-1},\,z)\\[.5cm]
\ds\phantom{[1-x+x_-(1-xz)]}=\ds\frac{1}{(x-1)}
\,\left(\frac{z(x
    z-1)^2(x_+-x_-)^2}{x(z-1)^2}\right)^n \,\,{\W}_n(z)
\end{array}\label{Wr1}
\eeq
of the quantum Wronskian relation \eqref{Wr}.
Here ${\W}_n(z)$ is related to $W(\q,n)$ in \eqref{Wr},
\beq
W(\q,n)=(-1)^n\,i\,\big[2\vt_1(\pi/3\,|\,\q)\big]^{2n+1}\, {\mathcal N}^2(\q,n)\, 
{\W}_n(z)\ .
\eeq
Equation \eqref{Wr1} is an algebraic identity valid for arbitrary
values of $x$. Expanding this identity around $x=z^{-1}$, one obtains 
\beq
{\W}_n(z)=-s_n(z)\, {\mathcal P}_n(z^{-1},z),\qquad n\ge0\ ,
\label{Wr-res}
\eeq
where $s_n(z)$ is defined by \eqref{gen19} (it coincides with the
Painlev\'e VI tau-function $s_n(z)=\tau_{n+1}(z,1/6)$ defined by 
\eqref{tau-rec} for $\xi=1/6$).
Similarly, expanding \eqref{Wr1} around $x=0$ and using
\eqref{Wr-res}, one obtains
\beq
{\mathcal P}_n(1,z)=4^n\,s_n(z),\qquad n\ge0\ .\label{P1}
\eeq
Interestingly the quantity ${\mathcal P}_n(z^{-1},z)$, entering
\eqref{Wr-res} is also determined by 
the Painlev\'e VI recurrence relation \eqref{tau-rec}.

\begin{conjD}The value ${\mathcal P}_n(z^{-1},z)$ is determined by 
recurrence relation \eqref{tau-rec} with $\xi=2/3$,
\beq
{\mathcal P}_n(z^{-1},z)=(-4/3)^n\, z^{-n}\, \tau_{n+2}(z,2/3),\qquad 
n\ge 0. \label{Pz}
\eeq
\end{conjD}
Combining the above formulae one obtains the following expression for
the quantum Wronskian,
\beq
{\W}_n(z)=-(-4/3)^n\,z^{-n} \tau_{n+1}(z,1/6)\,
\tau_{n+2}(z,2/3), \qquad n\ge0 \ .
\eeq

\section{Eigenvector results}\label{results}

To formulate our results for the eigenvectors \eqref{Psi-def} we need
to define an additional set of 
polynomials   $p_n(y)$ 
and $q_n(y)$, $n\in {\mathbb Z}$. In principle, these polynomials 
 can be defined by 
yet another recurrence relation of the Painlev\'e VI type (though more
complicated than \eqref{tau-rec}) which will be presented elsewhere. 
For our purposes here it is much simpler to 
define these new polynomials $p_n(y)$ 
and $q_n(y)$ as subfactors of already
introduced polynomials $s_n(z)$. We will do this by means of the 
Conjecture E, given below. Remind that $s_n(z)$, $n\in {\mathbb Z}$ are 
defined by 
Eq.\eqref{gen19} with the initial conditions $s_0(z)=s_1(z)\equiv1$. 

\begin{conjE} {\ }
\begin{enumerate}[(a)]
\item The polynomials $s_{2k+1}(y^2)$ factorize over the integers,  
\beq
s_{2k+1}(y^2)=s_{2k+1}(0)
\,p_k(y)\,p_k(-y),\qquad p_k(0)=1,\qquad k\in {\mathbb Z},\label{odd-fac}
\eeq
where $p_k(y)$ are polynomials in $y$ with integer coefficients, $\deg
p_k(y)=k(k+1)$, such that 
$p_k'(0)>0$,  $k\ge1$ and $p_k'(0)<0$, $k\le-2$, where $p_k'(y)=dp_k(y)/dy$
denotes the derivative in $y$.
Note that $p_{-1}(y)=p_{0}(y)\equiv1$. 
\item
the polynomials $p_k(y)$ possess the symmetry 
\beq
p_k(y)=\Big(\frac{1+3y}{2}\Big)^{k(k+1)}\,p_k\Big(\frac{1-y}{1+3y}\Big),\quad
k\in {\mathbb Z},\label{p-sym}
\eeq
\item
polynomials $s_{2k}(y^2)$ factorize over the integers, 
\beq
s_{2k}(y^2)=c_k\,(1+3y)^{k(k+1)}\,
p_{-k-1}\Big(\frac{y-1}{1+3y}\Big)\,q_{k-1}(y),\qquad k\in {\mathbb
  Z},\label{even-fac} 
\eeq
where $q_k(y)$ are polynomials in $y$ with integer coefficients, 
$\deg q_k(y)=k(k+1)$, $q_k(0)=1$ and 
\beq
c_k=2^{-k(k+2)},\quad k\ge0;\qquad 
c_k=2^{-k^2}\,(2/3)^{2k+1},\quad k<0\ .\label{ck-def}
\eeq
\item
polynomials $q_k(y)$ possess the symmetry
\beq
q_k(y)=\Big(\frac{1+3y}{2}\Big)^{k(k+1)}\,q_k\Big(\frac{y-1}{1+3y}\Big),\qquad
k\in {\mathbb Z},\label{q-sym}
\eeq
\end{enumerate}
\end{conjE} 
A few first polynomials $s_n(z)$, $p_n(y)$ and $q_n(y)$ is listed 
in Appendix~A. Let us mention one simple, but important corollary of
the above conjecture. The LHS of \eqref{even-fac} is an even functions
of the variable $y$. 
Combining this fact with the symmetry relation \eqref{p-sym},
one immediately deduce that $q_k(y)$ is also an even function, 
\beq
q_k(-y)=q_k(y),\qquad k\in\mathbb{Z}\ .\label{y-even}
\eeq
In Conjecture~1, given in the Introduction, we have stated 
an explicit expression \eqref{psi-norm} for the norm
$|\Psi_-|^2$ of the eigenvector $\Psi_-$ as a function of the
parameter $\zeta$ entering the Hamiltonian~\eqref{ham}.
Note, that using the factorization and symmetry properties 
(\ref{odd-fac})-(\ref{y-even}), one can show that the rescaled
norm 
\beq
{N}_n(\zeta)=(\zeta^2+3)^{-n(n+1)/2}\ |\Psi_-|^2\label{N-norm}
\eeq
is invariant with respect to the full $S_3$ symmetry group generated
by the substitutions \eqref{s3group}, which is a well expected
result. Further, is easy to see, that modulo a trivial numerical
factor, the expression for the norm remains unchanged upon the
replacement $n\to -n-1$, which corresponds to a negation of the length
of the chain, $N\to -N$. In other words, the norm is an even function
of the length of the chain. It would be interesting to understand a reason
of this phenomenon.

We are now ready to present further conjectures on the
properties of the eigenvectors.

\begin{conj} The component of the eigenvector $\Psi_-$
with one spin down is given by
\beq
{\ds \psi_{0\ldots001}=\frac{1}{N}\zeta^{n(n-1)/2}\,
\ov s_n(\zeta^{-2})},\qquad N=2n+1\ ,\label{con-one}
\eeq
where $\ov s_n(z)$ is defined by \eqref{r-coeff}.
\end{conj}
A few first polynomials $s_n(z)$ and $\ov s_n(z)$ is listed in Appendix~A.

\begin{conj}
The component of the eigenvector $\Psi_-$ with all spins down is given by
\beq
{\ds
  \psi_{11\ldots11}=\zeta^{n(n+1)/2}\,s_n(\zeta^{-2})},\qquad
N=2n+1\ ,\label{con-all} 
\eeq
where $s_n(z)$ is defined by \eqref{r-coeff}.
\end{conj}
It is interesting to note that to within a simple power of $\zeta$ 
the above two components of the eigenvector 
precisely coincide with the constant term and 
leading coefficients of the polynomial $\P_n(x,z)$, which is simply connected 
\eqref{Q1} with the corresponding eigenvalue of the $\mathsf{Q}$-operator.
We believe that this fact certainly deserves further studies.  

Finally, consider components of $\Psi_-$ 
with alternating (up and down) spins in the chain,
\beq
A_{n}(\zeta)=\Psi_{00101\ldots01},\quad \mbox{for odd $n$};\qquad\qquad
A_{n}(\zeta)=\Psi_{0101\ldots011}
\quad \mbox{for even $n$}\ .\label{alt}
\eeq
In the case $\zeta=0$ these are largest components of the
eigenvector. 

\begin{conj} The components of $\Psi_-$ with alternative spins are
  given by
\beq\begin{array}{rcl}
A_{2k}(\zeta)&=&2^{k(2-k)}\, (3+\zeta)^{k(k-1)}\,
\zeta^{k(k-1)}
p_{k-1}\Big(\frac{1-\zeta}{3+\zeta}\Big)\  q_{k-1}(\zeta^{-1})\\[.4cm]
A_{2k+1}(\zeta)&=& \ \  2^{-k^2} (3+\zeta)^{k(k+1)}\,
\zeta^{k(k-1)}\,
p_{k}\Big(\frac{1-\zeta}{3+\zeta}\Big)\  q_{k-1}(\zeta^{-1})
\end{array}\label{anzeta}
\eeq
\end{conj}
A few polynomials $p_n(y)$, $q_n(y)$ and $A_n(\zeta)$ is listed in Appendix~A.
As noted before, the case $\zeta=0$ corresponds to the 
$\mathsf{XXZ}$-model with $\Delta=-1/2$. 
It is known \cite{K02,SR2001,BdGN,dFZ05}, 
that in this particular case, 
the values of the components \eqref{alt}, normalized by \eqref{6vnorm},  
coincide with the number of alternating sign matrices
\beq
A_n(0)=A_n=\prod_{k=0}^{n-1}\frac{(3k+1)!}{(n+k)!}
\eeq
calculated in \cite{MRR83}. 
Using this result in \eqref{anzeta} one easily obtains 
for $n\ge0$,
\beq
p_n({\textstyle\frac{1}{3}})=\left(\frac{2}{3}\right)^{n(n+1)}\prod_{k=0}^n
\frac{(2k)!(6k+1)!}{(4k)!(4k+1)!}\ ,
\eeq
\beq
\zeta^{n(n+1)}q_n(\zeta^{-1})|_{\zeta=0}=
2^{-n-1}\prod_{k=0}^n
\frac{(2k+1)!(6k+4)!}{(4k+2)!(4k+3)!}\ .
\eeq 
Apparently one can derive these expressions directly from the definitions 
of the polynomials $p_n(y)$ and $q_n(y)$, given in Conjecture~E,
however, we postpone this to a future publication.

Finally, mention one amusing observation connected 
with the expressions \eqref{anzeta}. 
It is not difficult to analytically derive 
an asymptotic expansion $A^{(asymp)}_n(\zeta)$ 
which correctly reproduce first terms of the expansion of $A_n(\zeta)$
for small $\zeta$ up to the order $O(\zeta^{2n})$,
\beq
A_n(\zeta)=A^{(asymp)}_n(\zeta)+O(\zeta^{2n}),\qquad \zeta\to0 \ .
\eeq
Analytically continuing this asymptotic expansion to $n=0$,
\beq
A^{(asymp)}_n(\zeta)\Big|_{n=0}=1-\zeta^2-3\,\zeta^4-15\,\zeta^6-86\,\zeta^8
-534\,\zeta^{10}-3478\,\zeta^{12}-\ldots 
\eeq
and plugging its coefficients into Sloane's integer sequences database
(in a search for a discovery) we found that they only ``slightly'' 
mismatched numbers of lattice animals made of $n$ three-dimensional
cubes \cite{GB06}, which are $1,3,15,86,534,3481,\ldots$. Of course, it
would be extremely weird if they matched.

\section{Conclusion and outlook}
In this paper we have demonstrated that a 
particular anisotropic $\mathsf{XYZ}$-model, defined by \eqref{ham}
and \eqref{J-def}, is  
deeply related with the theory of Painlev\'e VI equation. We have
proposed exact expressions for the norm (Conjecture 1) and 
certain components of the ground state
eigenvectors (Conjectures 2,3,4). The results are expressed  
in terms of the tau-functions associated with the special elliptic
solutions of the Painlev\'e VI equation \cite{BM06a}. 

In this connection, it is useful to 
mention other celebrated appearances of  
Painlev\'e transcendents in mathematical physics. The most prominent examples 
include the two-dimensional Ising
model \cite{BMW}, the problem of isomonodromic deformations 
of the second order differential equations \cite{JMSM} and 
the field theory approach to dilute 
self-avoiding polymers on a cylinder
\cite{CV91,FS94,Zam94,TW96,Fen99}. 
The latter problem is connected with the massive sine-Gordon model 
at the supersymmetric point (where the ground state energy 
vanishes exactly due to supersymmetry). 
Our previous work \cite{BM05}
grew up from attempts to develop an alternative approach to this 
polymer problem based on the lattice theory.  It turns out that all
non-trivial information about dilute polymer loops is contained in
the ground state eigenvalues \cite{BM05} of the $\mathsf{Q}$-operator 
for the 8V-model on a periodic chain 
of an odd length, connected with 
the special $\mathsf{XYZ}$-model, considered in this paper. 
In \cite{BM05} we have found that these eigenvalues 
can be uniquely determined as certain polynomial solutions ${\mathcal
  P}_n(x,z)$ of the partial differential equation
\eqref{P-pde} (remind that the variable $x$ and $z$ are connected to 
the original spectral parameter $u$ and
the elliptic nome $\q$, respectively, see \eqref{x-def1}).
So far we have not ultimately understood the role of this equation in the  
Painlev\'e VI theory, but there is no doubt that there are profound
connections.  For example, 
one-variable specialization of
$\P_n(x,z)$ at particular values of $x$ (which remain polynomials in the
variable $z$) are connected with the tau-functions 
associated with the Picard solutions of the Painlev\'e VI equation
(see Eqs.\eqref{P1} and \eqref{Pz}).  
The same property is enjoyed also by
the coefficients in the expansion of $\P_n(x,z)$
in powers of $x$. Note, that namely these coefficients provide
``construction materials'' in the expression for the ground state
eigenvectors (see Eqs.\eqref{psi-norm},
\eqref{odd-fac}-\eqref{anzeta}). 
Most of our results are conjectures and it is, of course, desirable
to obtain their proofs. Another outstanding problem is an algebraic
construction of the $\mathsf{Q}$-matrix.   
As noted in \cite{McCoy1}, the method used in \cite{Bax72} for the
construction of the $\mathsf{Q}$-matrix cannot be executed in its full strength
for $\eta=\pi/3$, since some axillary $\Q$-matrix, $\Q_R(u)$, in
\cite{Bax72} is not invertible in the full $2^N$-dimensional space of
states of the model. Apparently, the construction of \cite{Bax72}
could be modified to resolve this difficulty. We hope to address this
question in the future. 

It is reasonable to expect that mathematical 
structures, similar to those described above
 (namely, the partial differential equations and Painlev\'e type recurrence relations),
 should manifest itself in
other problems, closely related to the 8V-model with $\eta=\pi/3$. The
most immediate candidate is the corresponding ``eight-vertex solid-on-solid'' 
(8VSOS) model which belongs to a rich variety of algebraic
constructions 
associated with the 8V-model \cite{Bax73abc}. Recently, Rosengren \cite{R09},
motivated by considerations of the 3-coloring problem \cite{Bax70c}, 
studied precisely this 8VSOS-model with $\eta=\pi/3$ in the case of 
the domain wall boundary conditions. In Eq.(8.11) of his paper \cite{R09} 
he introduced a set of two-variable polynomials (defined recursively), 
related to the
partition functions of the 8VSOS model on finite lattices.
A detailed inspection of these polynomials suggests that they satisfy a
 partial differential equation, which is completely analogous (though not
 identical) to our equation \eqref{P-pde}! This new differential
 equation is presented in Appendix~B. 

Furthermore, we found that some one-variable specialization of
Rosengren's polynomials, also defined in \cite{R09}, satisfy a
recurrence relation, which is extremely similar to the Painlev\'e IV
type recurrence relation \eqref{tau-rec} of this paper. This new
 relation is also presented in Appendix~B. As yet it 
is not written in a canonical form for Painlev\'e IV type relations,
however, we expect  
that it could be brought to such form by a suitable change of
variables. We also expect that this new recurrence relation for the
8VSOS-model can be connected with the Picard 
elliptic solutions of the Painlev\'e VI following the method of our
previous paper \cite{BM06a}. It seems it would be extremely interesting to
further compare our results with those of ref.\cite{R09}.

\section*{Acknowledgments}

The authors thank B.M.McCoy for valuable comments and 
H.Rosengren for sending us the preprint of his
recent paper \cite{R09} and interesting correspondence. 
After completion of this manuscript we received the preprint
\cite{RS09} on the same subject, but without   
essential overlaps with the present paper. 
We thank A.V.~Rasumov for sending us their preprint \cite{RS09}.

\app{Polynomials $\P_n(x,z)$, $s_n(z)$, $\ov s_n(z)$, $p_n(z)$ and $q_n(z)$.}

In this Appendix we present explicit expressions for the polynomials
$\P_n(x,z)$, $s_n(z)$, $\ov s_n(z)$, $p_n(z)$ and $q_n(z)$ for small values
of the their index $n$. 

The two-variable polynomials $\P_n(x,z)$, represented by
Eq.\eqref{P-def}, are defined as solutions of 
the $\mathsf{TQ}$-equation \eqref{TQ-alg}
normalized by \eqref{r-norm}. These polynomials can be efficiently calculated
from the differential equation \eqref{P-pde}, 
\beq
\begin{array}{l}
\P_0(x,z)=1, \quad \P_1(x,z)=x+3,\quad
\P_2(x,z)=x^2(1+z)+5x(1+3z)+10,\\[.4cm]
\P_3(x,z)=x^3(1+3z+4z^2)+7x^2(1+5z+18z^2)+7x(3+19z+18z^2)+35+21z,\\[.3cm]
\P_4(x,z)=x^4(1+6z+18z^2+30z^3+9z^4)+9x^3(1+8z+38z^2+152z^3+57z^4)\\[.3cm]
\phantom{\P_4(x,z)}
+18x^2(2+19z+111z^2+217z^3+99z^4)+12x(7+72z+171z^2+198z^3)\\[.3cm]
\phantom{\P_4(x,z)}+18(7+14z+11z^2).
\end{array}
\label{gen5}\eeq
The polynomials $s_n(z)$ and $\ov s_n(z)$, where $n\in{\mathbb Z}$, 
are defined by
Eq.\eqref{r-coeff} and the recurrence relation \eqref{tau-rec},
\eqref{tau-init}. For non-negative $n$ they coincide with the
coefficients of the highest and lowest powers of $x$ in $\P_n(x,z)$,
corresponding to $k=n$ and $k=0$ in the expansion
\eqref{P-def}, 
\beq
\begin{array}{l}
s_{-5}={\textstyle \frac{1}{256}}\,
(81+1215\,z+10206\,{z}^{2}+64638\,{z}^{3}+353565\,{z}^{4}+544563\,{z}^{
5}+352836\,{z}^{6}),\\[.3cm]
s_{-4}={\textstyle \frac{1}{64}}\,
(27+270\,z+1620\,{z}^{2}+7938\,{z}^{3}+3969\,{z}^{4}),\\[.3cm]
s_{-3}(z)={\textstyle \frac{1}{16}}\,(9+54\,z+225\,{z}^{2}),\\[.3cm]
s_{-2}(z)={\textstyle \frac{1}{4}}\,(3+9\,z),\\[.3cm]
s_{-1}(z)=s_0(z)=s_1(z)=1,\\[.3cm]
s_2(z)=1+z,\\[.3cm]
s_3(z)=1+3\,z+4\,{z}^{2},\\[.3cm]
s_4(z)=1+6\,z+18\,{z}^{2}+30\,{z}^{3}+9\,{z}^{4},\\[.3cm]
s_5(z)=
1+10\,z+51\,{z}^{2}+168\,{z}^{3}+355\,{z}^{4}+318\,{z}^{5}+121\,{z}^{6}.
\end{array}\eeq
and
\beq
\begin{array}{l}
\ov s_0(z)=1,\qquad \ov s_1(z)=3,\qquad \ov s_2(z)=10,\\[.3cm]
\ov s_3(z)=35+21\,z,\\[.3cm]
\ov s_4(z)=126+252\,z+198\,z^2,\\[.3cm]
\ov s_5(z)=462+1980\,z+3960\,z^2+4004\,z^3+858\,z^4,\\[.3cm]
\ov s_6(z)=1716+12870\,z+47190\,z^2+105820\,z^3+143520\,z^4+90558\,z^5+
24310\,z^6.\\[.3cm]
\end{array}\eeq
The polynomials $p_n(y)$ and $q_n(y)$ are defined by
the factorization relations \eqref{odd-fac} and \eqref{even-fac}
in Conjecture E,
\beq\begin{array}{l}
p_{-3}(y)=
1-3\,y+12\,{y}^{2}-30\,{y}^{3}+81\,{y}^{4}-63\,{y}^{5}+66\,{y}^{6},
\phantom{XXXXXXXXXXX}\\[.3cm]
p_{-2}(y)=1-2\,y+5\,{y}^{2},\\[.3cm]
p_{-1}(y)=p_0(y)=1,\\[.3cm]
p_1(y)=1+y+2\,{y}^{2},\\[.3cm]
p_2(y)=1+2\,y+7\,y^2+10\,y^3+21\,y^4+12\,y^5+11\,{y}^{6},\\[.3cm]
p_3(y)=
1+3\,y+15\,{y}^{2}+35\,{y}^{3}+105\,{y}^{4}+195\,{y}^{5}+435\,{y}^{6}\\[.3cm]
\phantom{p_3(y)a}
+555\,{y}^{7}+840\,{y}^{8}+710\,{y}^{9}+738\,{y}^{10}+294\,{y}^{11}+
170\,{y}^{12}.
\end{array}\eeq
and 
\beq
\begin{array}{l}
q_{-3}(y)=1+3\,{y}^{2}+39\,{y}^{4}+21\,{y}^{6},
\phantom{XXXXXXXXXXXXXXXXXX}\\[.3cm]
q_{-2}(y)=1+3\,y^2,\\[.3cm]
q_{-1}(y)=q_0(y)=1,\\[.3cm]
q_1(y)=1+3\,y^2,\\[.3cm]
q_2(y)=1+8\,y^2+29\,y^4+26\,{y}^{6},\\[.3cm]
q_3(y)=1+15\,{y}^{2}+112\,{y}^{4}+518\,{y}^{6}+1257\,{y}^{8}+1547\,{y}^{10}+
646\,{y}^{12}.
\end{array}\eeq

Finally, we list polynomials $A_n(\zeta)$ from the expressions
\eqref{anzeta} for the alternative spin components \eqref{alt},
\beq\begin{array}{l}
A_1(\zeta)=1,\quad A_2(\zeta)=2,\quad A_3(\zeta)=7+\zeta^2,\quad
A_4(\zeta)=2(3+\zeta^2)(7+\zeta^2),\>\nonumber\\[.3cm]
A_5(\zeta)=(3+\zeta^2)(143+99\zeta^2+13\zeta^4+\zeta^6),\\[.3cm]
A_6(\zeta)=2(26+29\zeta^2+8\zeta^4+\zeta^6)(143+99\zeta^2+13\zeta^4+\zeta^6),
\\[.3cm]
A_7(\zeta) = (26 + 29\zeta^2 + 8\zeta^4 + \zeta^6)\\[.3cm]
\phantom{A_8(\zeta) = }\times
(8398 + 14433\zeta^2 + 7665\zeta^4 + 
      2010\zeta^6 + 240\zeta^8 + 21\zeta^{10} + \zeta^{12}),\nonumber\\[.3cm]
A_8(\zeta) = 2(646 + 1547\zeta^2 + 1257\zeta^4 + 518\zeta^6 +
112\zeta^8 + 15\zeta^{10} + \zeta^{12})\\[.3cm]
\phantom{A_8(\zeta) = }
\times(8398 + 14433\zeta^2 + 7665\zeta^4 + 2010\zeta^6 + 240\zeta^8 + 
     21\zeta^{10} + \zeta^{12})\\[.3cm]
A_9(\zeta)=2(646+1547\,\zeta^2+1257\,\zeta^4+518\,\zeta^6+112\,\zeta^8+
      15\,\zeta^{10}+\zeta^{12})\\[.3cm]
\phantom{A_8(\zeta)=}\times(1411510 + 4598551\,\zeta^2 + 5518417\,\zeta^4 +
  3530124\,\zeta^6 + 1331064\,\zeta^8  + \\[.3cm]
\phantom{A_8(\zeta)=}
+ 327810\,\zeta^{10}
+53382\,\zeta^{12}+5820\,\zeta^{14}+506\,\zeta^{16}+31\,\zeta^{18}+
      \zeta^{20}).
\end{array}\eeq

\app{Comments on the 8VSOS-model}
Recently, Rosengren \cite{R09},
motivated by considerations of the 3-coloring problem \cite{Bax70c}, 
studied the 8VSOS-model with $\eta=\pi/3$ in the case of 
the domain wall boundary conditions. In Eq.(8.11) of his paper \cite{R09} 
he introduced a set of two-variable polynomials, related to the
partition functions of the 8VSOS model on finite lattices. Here denote
these polynomials as $P^{(SOS)}_n(t,s)$, adding the
superscript ``SOS'' to indicate their relevance to the 8VSOS-model\footnote{%
Here we use the variables $t$ and $s$ instead of  
$x$ and $\zeta$ used in \cite{R09}. These variables 
are similar, {\em but not identical}, to our variables $x$ and $z$ in
\eqref{x-def1}. In particular, the variable $t$ 
(which corresponds to $x$ in \cite{R09}) is also connected with 
the spectral parameter,  while the variable
$s$ (denoted as $\zeta$ in \cite{R09}) is related to the elliptic nome $\q$.}.
A detailed inspection of these polynomials suggests that 
\begin{conj} 
The polynomials
  $P^{(SOS)}_n(t,s)$, for even values of $n=0,2,4\ldots$, 
are uniquely determined (up to a numerical normalization) 
by the  the following
partial differential equation in the variables $x$ and $s$,
\beq
\Big\{A^{(SOS)}(t,s)\,\partial_t^2 +B^{(SOS)}_n(t,s)\,
\partial_t +C^{(SOS)}_n(t,s)\, +
T^{(SOS)}(t,s)\,{\ds\partial_s}\Big\}\ P^{(SOS)}_n(t,s)=0,
\label{sos-pde}
\eeq
where
\beq\begin{array}{rl}
A^{(SOS)}(t,s)&=
2\, t\left( 1-t \right) \left( 1+2\,s-t \right)  \left( s+2\,{s}^{2}-
2\,t-st \right)  \left(2\,t+st-s \right),\\[.4cm] 
B_n^{(SOS)}(t,s)&=
-4\, \left( 2+s \right) ^{2}{t}^{4}-4\, \left( 2+s \right)  \left( -2
\,{s}^{2}+{s}^{2}n+3\,sn-5\,s-3+n \right) {t}^{3}\\[.4cm]
& +\left( 40\,sn+8\,{s
}^{4}n+40\,{s}^{3}n+60\,{s}^{2}n+8\,n-8\,{s}^{4}-60\,{s}^{2}-8-44\,{s}
^{3}-36\,s \right) {t}^{2}\\[.4cm]
&-4\, \left( 1+2\,s \right) s \left( -{s}^{2}
+3\,{s}^{2}n+5\,sn-3\,s+3\,n-2 \right) t+4\,n{s}^{2} \left( 1+2\,s
 \right) ^{2},\\[.4cm]
C_n^{(SOS)}(t,s)&=
2\,n \left( 2+s \right) ^{2} \left( 1+n \right) {t}^{3}-4\,n \left( 2+
s \right)  \left( 1+s \right) ^{2} \left( 1+n \right) {t}^{2}\\[.4cm]
&+ns
 \left( 8\,{s}^{3}+4\,{s}^{3}n+26\,{s}^{2}+15\,{s}^{2}n+18\,s+18\,sn+2
+5\,n \right) t\\[.4cm]
&-4\,n \left( 1+2\,s \right) s \left( s+sn+1 \right),\\[.4cm] 
T^{(SOS)}(t,s)&=4\, \left( 1-s^2 \right) s \left( 2+s \right) 
 \left( 1+2\,s \right) t\ .
\end{array}
\eeq
\end{conj}
A similar differential equation exists for odd values of $n$.
Obviously, the above property is a counterpart of the partial differential 
equation \eqref{P-pde} in the main text of this paper. 

Next, 
define one-variable polynomials\footnote{%
These polynomials are simply related to those 
introduced in \cite{R09}, 
$$
p^{(SOS)}_n(s)=
\big(1+2s\big)^{[\frac{n^2}{4}]}
\big(1+\frac{s}{2}\big)^{[\frac{(n-1)^2}{4}]}\ p^{(R)}_n(s)
$$  
where ${p}^{(R)}_n(s)$ are defined by the first unnumbered
equation after the Proposition~3.12 in \cite{R09}.
We thanks H.Rosengren for sending us the modified
definition \eqref{sos-pol}.} 
\beq\label{sos-pol}
p^{(SOS)}_n(s)=
\Big(1+2s\Big)^{\big[\frac{n^2}{4}\big]-\big[\frac{n}{2}\big]}
\Big(1+\frac{s}{2}\Big)^{\big[\frac{(n-1)^2}{4}\big]}\ P_n^{(SOS)}(1+2s,s)
\eeq  
where $[x]$ denotes the integer part of $x$. We suggest that 
\begin{conj}
The polynomials  $p^{(SOS)}_n(s)$ satisfy the following recurrence
relations 
\beq\begin{array}{l}
s(s-1)^2(s+2)(2s+1)\partial^2_s \log p_n^{(SOS)}(s)
+2(s-1)(s^3-3s^2-6s-1)
\partial_s\log p_n^{(SOS)}(s)\\[.3cm]
\phantom{XXXXXX}\ds
-4(2n+1)(2n+3)\frac{p_{n+1}^{(SOS)}(s)p_{n-1}^{(SOS)}(s)}
{(p_n^{(SOS)}(s))^2}\\[.5cm]
\phantom{XXXXXX}
+(22n^2+35n+18)\,s^2+(46n^2+98n+42)\,s+13n^2+29n+12=0
\end{array}\label{sos2}
\eeq
with the initial condition $p_0(s)=1$, $p_1(s)=1+3s$. 
\end{conj}
The structure of the relation \eqref{sos2} is very similar
to that of the recurrence relation \eqref{tau-rec} for the tau-functions
of Painlev\'e VI equation. It should be noted that 
Eq.\eqref{sos2} is not written in a canonical 
form for such recurrence relations. Nevertheless, we expect 
that it could be brought to such form by a suitable change of
variables. We also expect that \eqref{sos2} can be connected with the Picard 
elliptic solutions of the Painlev\'e VI following the method of our
previous paper \cite{BM06a}. 

\newcommand\oneletter[1]{#1}

%\bibliography{refs225}

\begin{thebibliography}{10}

\bibitem{BM06a}
Bazhanov, V.~V. and Mangazeev, V.~V.
\newblock Eight vertex model and {P}ainlev\'e {V}{I}.
\newblock J. Phys. A {\bf 39} (2006) 12235--12243.
\newblock arXiv:hep-th/0602122.

\bibitem{Bax72}
Baxter, R.~J.
\newblock Partition function of the eight-vertex lattice model.
\newblock Ann. Physics {\bf 70} (1972) 193--228.

\bibitem{Bax72b}
Baxter, R.~J.
\newblock One-dimensional anisotropic {H}eisenberg chain.
\newblock Ann. Phys. {\bf 70} (1972) 323--337.

\bibitem{Str01a}
Stroganov, Y.
\newblock The importance of being odd.
\newblock J. Phys. A {\bf 34} (2001) L179--L185.

\bibitem{BM05}
Bazhanov, V.~V. and Mangazeev, V.~V.
\newblock Eight-vertex model and non-stationary Lame equation.
\newblock J. Phys. A {\bf 38} (2005) L145--153.
\newblock arXiv/hep-th/0411094.

\bibitem{Str01c}
Stroganov, Y.
\newblock The {$8$}-vertex model with a special value of the crossing parameter
  and the related {$XYZ$} spin chain.
\newblock In {\em Integrable structures of exactly solvable two-dimensional
  models of quantum field theory (Kiev, 2000)}, volume~35 of {\em NATO Sci.
  Ser. II Math. Phys. Chem.}, pages 315--319. Kluwer Acad. Publ., Dordrecht,
  2001.

\bibitem{SR2001}
Razumov, A.~V. and Stroganov, Y.~G.
\newblock Spin chains and combinatorics.
\newblock J. Phys. A {\bf 34} (2001) 3185--3190.

\bibitem{Bax73abc}
Baxter, R.~J.
\newblock Eight-vertex model in lattice statistics and one-dimensional
  anisotropic {H}eisenberg chain. {I}. {S}ome fundamentaleigenvectors. {I}{I}.
  {E}quivalence to a generalized ice-type lattice model. {I}{I}{I}.
  {E}igenvectors of the transfer matrix and {H}amiltonian.
\newblock Ann. Phys. {\bf 76} (1973) 1--24, 25--47, 48--71.

\bibitem{R09}
Rosengren, H.
\newblock The three-colour model with domain wall boundary conditions, 2009.
\newblock {\tt{arXiv:0911.0561}}.

\bibitem{Bax70c}
Baxter, R.~J.
\newblock Three-colorings of the square lattice: a hard squares model.
\newblock J. Math. Phys. {\bf 11} (1970) 3116--3124.

\bibitem{WW}
Whittaker, E. and Watson, G.
\newblock {\em A course of modern analysis}.
\newblock ``Cambridge University Press'', Cambridge, 1996.

\bibitem{BLZ97a}
Bazhanov, V.~V., Lukyanov, S.~L., and Zamolodchikov, A.~B.
\newblock Integrable structure of conformal field theory. {I}{I}. ${
  Q}$-operator and {D}{D}{V} equation.
\newblock Comm. Math. Phys. {\bf 190} (1997) 247--278.
\newblock [{\tt hep-th/9604044}].

\bibitem{KLWZ97}
Krichever, I., Lipan, O., Wiegmann, P., and Zabrodin, A.
\newblock Quantum integrable models and discrete classical {H}irota equations.
\newblock Comm. Math. Phys. {\bf 188} (1997) 267--304.
\newblock [{\tt hep-th/9604080}].

\bibitem{McCoy2}
Fabricius, K. and McCoy, B.~M.
\newblock New Developments in the Eight Vertex Model II. Chains of odd length.
\newblock (2004).
\newblock cond-mat/0410113.

\bibitem{BM07}
Bazhanov, V.~V. and Mangazeev, V.~V.
\newblock Analytic theory of the eight-vertex model.
\newblock Nucl. Phys. B {\bf 775} (2007) 225--282.

\bibitem{EK94}
Etingof, P.~I. and Kirillov, Jr., A.~A.
\newblock Representations of affine {L}ie algebras, parabolic differential
  equations, and {L}am\'e functions.
\newblock Duke Math. J. {\bf 74} (1994) 585--614.

\bibitem{Fat09}
Fateev, V.~A., Litvinov, A.~V., Neveu, A., and Onofri, E.
\newblock Differential equation for four-point correlation function in
  Liouville field theory and elliptic four-point conformal blocks.
\newblock J. Phys. A {\bf 42} (2009) 304011.

\bibitem{K02}
Kuperberg, G.
\newblock Symmetry classes of alternating-sign matrices under one roof.
\newblock Ann. of math. {\bf 2} (2002) 156.

\bibitem{BdGN}
Batchelor, M.~T., de~Gier, J., and Nienhuis, B.
\newblock The quantum symmetric {$XXZ$} chain at {$\Delta=-\frac 12$},
  alternating-sign matrices and plane partitions.
\newblock J. Phys. A {\bf 34} (2001) L265--L270.

\bibitem{dFZ05}
Di~Francesco, P. and Zinn-Justin, P.
\newblock The quantum {K}nizhnik-{Z}amolodchikov equation, generalized
  {R}azumov-{S}troganov sum rules and extended {J}oseph polynomials.
\newblock J. Phys. A {\bf 38} (2005) L815--L822.

\bibitem{MRR83}
Mills, W.~H., Robbins, D.~P., and Rumsey, Jr., H.
\newblock Alternating sign matrices and descending plane partitions.
\newblock J. Combin. Theory Ser. A {\bf 34} (1983) 340--359.

\bibitem{GB06}
Aleksandrowicz, G. and Barequet, G.
\newblock Counting d-Dimensional Polycubes and Nonrectangular Planar
  Polyominoes.
\newblock Lect. Notes in Comp. Sci. {\bf 4112/2006} (2006) 418--427.

\bibitem{BMW}
Barouch, E., McCoy, B.~M., and Wu, T.~T.
\newblock Zero-field susceptibility of the two-dimensional Ising model near
  $T_c$.
\newblock Phys. Rev. Lett. {\bf 31} (1973) 1409--1411.

\bibitem{JMSM}
Jimbo, M., Miwa, T., Sato, M., and M{\^o}ri, Y.
\newblock Holonomic quantum fields. {T}he unanticipated link between
  deformation theory of differential equations and quantum fields.
\newblock In {\em Mathematical problems in theoretical physics (Proc. Internat.
  Conf. Math. Phys., Lausanne, 1979)}, volume 116 of {\em Lecture Notes in
  Phys.}, pages 119--142. Springer, Berlin, 1980.

\bibitem{CV91}
Cecotti, S. and Vafa, C.
\newblock Topological--anti-topological fusion.
\newblock Nuclear Phys. B {\bf 367} (1991) 359--461.

\bibitem{FS94}
Cecotti, S., Fendley, P., Saleur, H., Intriligator, K., and Vafa, C.
\newblock A new supersymmetric index.
\newblock Nucl. Phys. B {\bf 386} (1992) 405--452.

\bibitem{Zam94}
Zamolodchikov, {\relax Al}.~B.
\newblock Painleve III and $2$D polymers.
\newblock Nuclear Phys. B {\bf 432} (1994) 427--456.

\bibitem{TW96}
Tracy, C.~A. and Widom, H.
\newblock Proofs of two conjectures related to the thermodynamic Bethe ansatz.
\newblock Comm. Math. Phys. {\bf 179} (1996) 1--9.

\bibitem{Fen99}
Fendley, P.
\newblock Airy functions in the thermodynamic {B}ethe ansatz.
\newblock Lett. Math. Phys. {\bf 49} (1999) 229--233.

\bibitem{McCoy1}
Fabricius, K. and McCoy, B.~M.
\newblock New developments in the eight vertex model.
\newblock J. Statist. Phys. {\bf 111} (2003) 323--337.

\bibitem{RS09}
Razumov, A.~V. and Stroganov, Y.~G.
\newblock A possible combinatorial point for XYZ-spin chain, 2009.
\newblock arXiv:0911.5030.

\end{thebibliography}
%\bibliographystyle{vvb-bibstyle}
%\bibliographystyle{alpha}
\end{document}